# Unified Performance Analysis of Mixed Line of Sight RF-FSO Fixed Gain Dual-Hop Transmission Systems

Emna Zedini, Imran Shafique Ansari, and Mohamed-Slim Alouini

Computer, Electrical, and Mathematical Sciences and Engineering (CEMSE) Division,
King Abdullah University of Science and Technology (KAUST),
Thuwal, Makkah Province, Saudi Arabia.
Emails: {emna.zedini, imran.ansari, slim.alouini}@kaust.edu.sa

*Abstract*—In this work, we carry out a unified performance analysis of a dual-hop fixed gain relay system over asymmetric links composed of both radio-frequency (RF) and unified free-space optics (FSO) under the effect of pointing errors. The RF link is modeled by the Nakagami-$m$ fading channel and the FSO link by the Gamma-Gamma fading channel subject to both types of detection techniques (i.e. heterodyne detection and intensity modulation with direct detection (IM/DD)). In particular, we derive new unified closed-form expressions for the cumulative distribution function, the probability density function, the moment generation function, and the moments of the end-to-end signal-to-noise ratio of these systems in terms of the Meijer's G function. Based on these formulas, we offer exact closed-form expressions for the outage probability, the higher-order amount of fading, and the average bit-error rate of a variety of binary modulations in terms of the Meijer's G function. Further, an exact closed-form expression for the end-to-end ergodic capacity for the Nakagami-$m$-unified FSO relay links is derived in terms of the bivariate G function. All the given results are verified via Computer-based Monte-Carlo simulations.

## I. Introduction

The demand for higher data rate raises the question of spectrum availability. In many wireless communications systems, data capacity has been enhanced by increasing the spectral efficiency by means of signal processing techniques and advanced modulation schemes. However, reaching rates of 10 Gbps or more is quite challenging due to the major limitation factor, scarce spectrum resources. For future communication systems, more spectral resources are mandatory. Of the many popular solutions, free-space optical (FSO) communication systems have gained significant research attention as effective means of transferring data at high rates over short distances mainly because they can provide not only higher capacity but also wider bandwidth relative to the traditional radio frequency (RF) transmission. Additionally, FSO links offer

This work was supported in part by a grant from King Abdulaziz City of Sciences and Technology (KACST).

a high security level, a license-free spectrum, and hence a cost effective solution compared to the RF communication systems. These advantages of FSO communication systems potentially help solving the issues that the RF communication systems face due to the expensive and scarce spectrum [1]–[6]. However, a number of technical challenges need to be overcome or better understood. In fact, contrary to RF links, the major severe limiting factor in FSO communications is its high vulnerability to the atmospheric turbulence conditions [2]. Fog, snow, and rain may cause a severe degradation in the overall performance. Another dominant factor affecting the reliability of FSO channels is building sway caused by thermal expansion, wind loads, and weak earthquakes [7], [8]. This phenomenon leads to a misalignment between the transmitter and the receiver defined as pointing error, which is a serious problem that degrades the channel performance [1], [9]. It is worthy to mention that the main type of detection technique in FSO systems is IM/DD. Coherent modulation is recently employed as an alternative detection approach. Despite of the complexity of implementing coherent receivers relative to IM/DD systems, heterodyne detection that belongs to coherent mode offers better performance in overcoming thermal noise effects [10], [11].

Relaying technique has gained an enormous interest due to its advantages including not only wider and energy-efficient coverage but also increased capacity in the wireless communication systems. Recently, several efforts have been conducted to investigate the relay system performance under various fading conditions [12]–[15]. Moreover, literature regarding the asymmetric relay networks based on both RF as well as FSO characteristics includes [16]–[19]. In [16], the performance analysis of asymmetric dual-hop RF-FSO relay system is presented. In [18], the performance analysis of a dual-hop variable gain relay RF-unified FSO transmission system subject to pointing errors is presented. [17] investigates the performance of a dual-branch fixed gain relay RF-FSO transmission system under the effect of pointing errors and subject to both types of detection techniques (i.e. IM/DD as well as heterodyne detection). However, the results presented in [17] were derived under the assumption of a non light of sight (NLOS) Rayleigh fading in the RF link, and as such does not cover the case when a line of sight (LOS) component is present between the source and relay. Since the Rician and Nakagami-$m$ fading models are more appropriate for propagation environments in LOS communications [20], in this work, we extend the model presented in [17] to study for the first time the performance of asymmetric LOS RF-FSO dual-hop fixed gain relay transmission systems with mixed Nakagami-$m$ RF-unified FSO links, which is a non-trivial contribution. More specifically, the FSO link is assumed to be operating over unified Gamma-Gamma fading environment [3], [4] under the effect of pointing errors, and the RF link over Nakagami-$m$ fading that includes the Rayleigh fading as a special case. In this context and in our performance analysis study, we used the finite series representation of the incomplete Gamma function along with the binomial expansion to derive unified exact closed-form expressions for the cumulative distribution function (CDF), the probability density function (PDF), the moment generating function (MGF), the moments, the higher-order amount of fading (AF), the outage probability (OP), the bit-error rate (BER) of a binary modulation schemes in terms of the Meijer's G function. Additionally, we present the ergodic capacity in terms of the bivariate G function. Further, we introduce the asymptotic expressions for all the expressions derived earlier in terms of the Meijer's G function at high signal-to-noise ratio (SNR) regime in terms of simple elementary functions by utilizing Meijer's G function expansion.

The remainder of the paper is organized as follows. In Section II, the system model and channel model for fixed Nakagami-$m$-unified FSO relay scheme is introduced. Exact closed-form results to characterize Nakagami-$m$-unified FSO relay including the PDF, the CDF, the MGF, the moments, the AF, the OP, the BER, and the ergodic capacity followed by the asymptotic expressions are presented in Sections III and IV. The derived analytical expressions in the previous sections are numerically evaluated, illustrated, and interpreted in Section V. Finally, we review our main results and we draw some conclusions in Section VI.

## II. CHANNEL AND SYSTEM MODELS

We consider an asymmetric dual-hop relaying system where the source node S and the destination node D are communicating through an intermediate relay node R. The RF point-to-point propagation link (i.e. S-R link) is assumed to follow a Nakagami-$m$ distribution. On the other hand, we assume the second FSO link (i.e. R-D link) experiences unified Gamma-Gamma fading with pointing error impairments. In the fixed gain relaying scheme, the end-to-end SNR can be expressed as [12]

$$\gamma = \frac{\gamma_1 \gamma_2}{\gamma_2 + C}, \quad (1)$$

where $\gamma_1$ denotes the SNR of the S-R hop, $\gamma_2$ represents the SNR of the R-D hop, and $C$ stands for a fixed relay gain [12], [16].

In this paper, we assume that the RF S-R link experiences Nakagami-$m$ fading distribution with the PDF in [20]

$$f_{\gamma_1}(\gamma_1) = \left(\frac{m}{\Omega}\right)^m \frac{\gamma_1^{m-1}}{\Gamma(m)} \exp\left(-\frac{m}{\Omega}\gamma_1\right), \quad (2)$$

where $m$ is the Nakagami-$m$ fading parameter ($m \geq \frac{1}{2}$), $\Gamma(\cdot)$ is the Gamma function as defined in [21, Eq.(8.310)], and $\Omega$ represents the average fading power, i.e. $\Omega = \mathbb{E}_{\gamma_1}[\gamma_1]$ with $\mathbb{E}$ denoting the expectation operator. It is important to note that the PDF in (2) includes the Rayleigh distribution ($m = 1$) as a special case.

The FSO R-D link is assumed to follow a unified Gamma Gamma fading distribution with pointing error impairments for which the PDF of the SNR is given by [22]

$$f_{\gamma_2}(\gamma_2) = \frac{\xi^2}{r \, \gamma_2 \, \Gamma(\alpha)\Gamma(\beta)} \, \mathrm{G}_{1,3}^{3,0}\left[h\,\alpha\,\beta\left(\frac{\gamma_2}{\mu_r}\right)^{\frac{1}{r}} \left| \begin{matrix} \xi^2 + 1 \\ \xi^2, \alpha, \beta \end{matrix} \right.\right], \quad (3)$$

where $r$ is the parameter specifying the detection technique type (i.e. $r = 1$ accounts for heterodyne detection and $r = 2$ represents IM/DD), $h = \frac{\xi^2}{\xi^2+1}$, $\xi$ denotes the ratio between the equivalent beam radius at the receiver and the pointing error displacement standard deviation (jitter) at the receiver [1], [23] (i.e. for negligible pointing errors, $\xi \to \infty$), $\alpha$ and $\beta$ are the the fading/scintillation parameters related to the atmospheric turbulence conditions with small values of these two parameters pointing to severe fading conditions [4], [9], $\mathrm{G}^{\cdot,\cdot}_{\cdot,\cdot}(\cdot)$ is the Meijer's G function as defined in [21, Eq.(9.301)], and $\mu_r$ standing for the average electrical SNR. More specifically, for $\mu_r$, when $r = 1$, $\mu_1 = \mu_{\text{heterodyne}} = \mathbb{E}[\gamma_2] = \overline{\gamma_2}$ and when $r = 2$, $\mu_2 = \mu_{\text{IM/DD}} = \overline{\gamma_2}\,\alpha\,\beta\,\xi^2\,(\xi^2+2)/[(\alpha+1)(\beta+1)(\xi^2+1)^2]$ [22].

## III. STATISTICAL CHARACTERISTICS

### A. Cumulative Distribution Function

The CDF of $\gamma$ is given by

$$F_\gamma(\gamma) = \Pr\left[\frac{\gamma_1\,\gamma_2}{\gamma_2 + C} < \gamma\right], \quad (4)$$

which can be expressed as

$$F_\gamma(\gamma) = \int_0^\infty \Pr\left[\frac{\gamma_1\,\gamma_2}{\gamma_2 + C} < \gamma | \gamma_2\right] f_{\gamma_2}(\gamma_2)\,\mathrm{d}\gamma_2$$
$$= 1 - \frac{\xi^2}{r\,\Gamma(\alpha)\,\Gamma(\beta)\,\Gamma(m)} \int_0^\infty \frac{1}{\gamma_2} \Gamma\left(m, \frac{m\,\gamma\,(\gamma_2+C)}{\Omega\,\gamma_2}\right)$$
$$\times \mathrm{G}_{1,3}^{3,0}\left[h\,\alpha\,\beta\left(\frac{\gamma_2}{\mu_r}\right)^{\frac{1}{r}} \left| \begin{matrix} \xi^2 + 1 \\ \xi^2, \alpha, \beta \end{matrix} \right.\right] d\gamma_2. \quad (5)$$

To the best of the authors knowledge, the solution to the integral in (5) is not available in exact closed-form nor in terms of the extended generalized bivariate Meijer's G function (EGBMGF) because of the shift in the incomplete Gamma function. Therefore, we utilize the finite series representation of the incomplete Gamma function in [21, Eq.(8.352.7)] to rewrite $\Gamma\left(m, \frac{m\,\gamma\,(\gamma_2+C)}{\Omega\,\gamma_2}\right)$ as

$(m-1)!\exp(-\frac{m\gamma}{\Omega})\exp(-\frac{m\,C\,\gamma}{\Omega\,\gamma_2})\sum_{k=0}^{m-1}\frac{1}{k!}\left(\frac{m\,\gamma}{\Omega}\right)^k\left(1+\frac{C}{\gamma_2}\right)^k$.
Since the summation is upper limited by $m$, our results are limited to the case of Nakagami-$m$ with integer values of $m$. Further using the binomial expansion in [21, Eq.(1.111)], $\left(1+\frac{C}{\gamma_2}\right)^k$ can be expressed as $\sum_{j=0}^{k}\binom{k}{j}\left(\frac{C}{\gamma_2}\right)^j$. Now, along with the above modifications, we apply [24,

Eq.(07.34.21.0088.01)] and some mathematical manipulations to get the CDF of $\gamma$ as

$$F_\gamma(\gamma) = 1 - A \exp\left(-\frac{m\gamma}{\Omega}\right) \sum_{k=0}^{m-1} \sum_{j=0}^{k} \frac{1}{j!(k-j)!} \left(\frac{m\gamma}{\Omega}\right)^{k-j}$$
$$\times G_{r,3r+1}^{3r+1,0}\left[\frac{BmC\gamma}{\mu_r \Omega} \bigg| \begin{matrix} \kappa_1 \\ \kappa_2 \end{matrix}\right], \quad (6)$$

where $A = \frac{r^{\alpha+\beta-2}\xi^2}{(2\pi)^{r-1}\Gamma(\alpha)\Gamma(\beta)}$, $B = \frac{(h\alpha\beta)^r}{r^{2r}}$, $\kappa_1 = \frac{\xi^2+1}{r},\ldots,\frac{\xi^2+r}{r}$ comprises $r$ terms, and $\kappa_2 = \frac{\xi^2}{r},\ldots,\frac{\xi^2+r-1}{r},\frac{\alpha}{r},\ldots,\frac{\alpha+r-1}{r},\frac{\beta}{r},\ldots,\frac{\beta+r-1}{r},j$ comprises $3r+1$ terms. For $m=1$, as a special case, the CDF in (6) is in agreement with the CDF of the hybrid Rayleigh/FSO fixed gain dual hop transmission systems with pointing errors presented in [17, Eq.(2)]. The arguments of the Meijer's G function in (6) can be inverted using [25, Eq.(6.2.2)]. Then, by applying [22, Eq.(26)], the asymptotic expression of the CDF at **high SNR** can be derived in terms of basic elementary functions as

$$F_\gamma(\gamma) \underset{\mu_r \gg 1}{\approx} 1 - A \exp\left(-\frac{m\gamma}{\Omega}\right) \sum_{k=0}^{m-1} \sum_{j=0}^{k} \frac{1}{j!(k-j)!} \left(\frac{m\gamma}{\Omega}\right)^{k-j}$$
$$\times \sum_{i=1}^{3r+1} \left(\frac{\Omega \mu_r}{BmC\gamma}\right)^{-\kappa_{2,i}} \frac{\prod_{l=1;l\neq i}^{3r+1} \Gamma(\kappa_{2,l} - \kappa_{2,i})}{\prod_{l=1}^{r} \Gamma(\kappa_{1,l} - \kappa_{2,i})}, \quad (7)$$

where $\kappa_{u,v}$ stands for the $v^{\text{th}}$-term of $\kappa_u$. This asymptotic expression for the CDF in (7) can be further expressed via only one dominant term, $j$, that represents the $(3r+1)^{\text{th}}$-term in $\kappa_2$.

### B. Probability Density Function

The PDF of $\gamma$ can be obtained by differentiating (6) with respect to $\gamma$. Therefore, utilizing the product rule then applying [24, Eq.(07.34.20.0001.01)], we get after some algebraic manipulations the PDF in exact closed-form in terms of the Meijer's G functions as

$$f_\gamma(\gamma) = A \exp\left(-\frac{m\gamma}{\Omega}\right) \sum_{k=0}^{m-1} \sum_{j=0}^{k} \frac{1}{j!(k-j)!} \left(\frac{m\gamma}{\Omega}\right)^{k-j}$$
$$\times \left\{ \left(\frac{m}{\Omega} - \frac{k-j}{\gamma}\right) G_{r,3r+1}^{3r+1,0}\left[\frac{BmC\gamma}{\mu_r \Omega} \bigg| \begin{matrix} \kappa_1 \\ \kappa_2 \end{matrix}\right] \right.$$
$$\left. - \frac{1}{\gamma} G_{r+1,3r+2}^{3r+1,1}\left[\frac{BmC\gamma}{\mu_r \Omega} \bigg| \begin{matrix} 0, \kappa_1 \\ \kappa_2, 1 \end{matrix}\right] \right\}. \quad (8)$$

For $m = 1$, as a special case, the PDF in (8) is in a perfect agreement with the PDF in [17, Eq.(3)].

### C. Moment Generating Function

It is well known that the MGF is defined as $\mathcal{M}_\gamma(s) = \mathbb{E}[e^{-\gamma s}]$. Using integration by parts, the MGF can be expressed in terms of CDF as

$$\mathcal{M}_\gamma(s) = s \int_0^\infty e^{-\gamma s} F_\gamma(\gamma) \, d\gamma. \quad (9)$$

Placing (6) into (9) and utilizing [24, Eq.(07.34.21.0088.01)], the MGF of $\gamma$ can be presented as

$$\mathcal{M}_\gamma(s) = 1 - s A \sum_{k=0}^{m-1} \sum_{j=0}^{k} \frac{\left(\frac{m}{\Omega}\right)^{k-j}}{j!(k-j)!} \left(s + \frac{m}{\Omega}\right)^{j-k-1}$$
$$\times G_{r+1,3r+1}^{3r+1,1}\left[\frac{BmC}{\mu_r(\Omega s + m)} \bigg| \begin{matrix} j-k, \kappa_1 \\ \kappa_2 \end{matrix}\right]. \quad (10)$$

When $m = 1$, as a special case, the MGF in (10) can be easily shown to be equal to [17, Eq.(6)]. Similar to the CDF, the asymptotic expansion of the MGF **high SNR** can be determined as

$$\mathcal{M}_\gamma(s) \underset{\mu_r \gg 1}{\approx} 1 - s A \sum_{k=0}^{m-1} \sum_{j=0}^{k} \frac{\left(\frac{m}{\Omega}\right)^{k-j}}{j!(k-j)!} \left(s + \frac{m}{\Omega}\right)^{j-k-1}$$
$$\times \sum_{i=1}^{3r+1} \left(\frac{\mu_r(\Omega s + m)}{BmC}\right)^{-\kappa_{2,i}}$$
$$\times \frac{\prod_{l=1;l\neq i}^{3r+1} \Gamma(\kappa_{2,l} - \kappa_{2,i})\Gamma(1 + \kappa_{2,i} - j + k)}{\prod_{l=2}^{r+1} \Gamma(\kappa_{1,l} - \kappa_{2,i})}, \quad (11)$$

and can be further expressed via only the dominant term, $j$, which is the $(3r+1)^{\text{th}}$-term in $\kappa_2$.

### D. Moments

The moments specified as $\mathbb{E}[\gamma^n]$ can be derived in terms of the complementary CDF (CCDF) $F_\gamma^c(\gamma) = 1 - F_\gamma(\gamma)$, via integration by parts, as

$$\mathbb{E}[\gamma^n] = n \int_0^\infty \gamma^{n-1} F_\gamma^c(\gamma) \, d\gamma. \quad (12)$$

Placing (6) into (12) and applying [24, Eq.(07.34.21.0088.01)], the moments reduce to

$$\mathbb{E}[\gamma^n] = n A \left(\frac{\Omega}{m}\right)^n \sum_{k=0}^{m-1} \sum_{j=0}^{k} \frac{1}{j!(k-j)!}$$
$$\times G_{r+1,3r+1}^{3r+1,1}\left[\frac{BC}{\mu_r} \bigg| \begin{matrix} 1-k+j-n, \kappa_1 \\ \kappa_2 \end{matrix}\right]. \quad (13)$$

For $m = 1$, as a special case, the moments in (13) can be easily shown to agree with [17, Eq.(8)]. It is important to mention that the moments are exploited to derive the expressions of the higher-order amount of fading in the next section.

## IV. APPLICATIONS TO THE PERFORMANCE OF ASYMMETRIC NAKAGAMI-$m$-UNIFIED FSO RELAY TRANSMISSION SYSTEMS WITH FIXED GAIN RELAY

### A. Outage Probability

The OP is an important measure for the performance of a wireless communication system. An outage of the communication system is encountered when the instantaneous output SNR $\gamma$ falls below a predetermined threshold $\gamma_{th}$. Setting $\gamma = \gamma_{th}$ in (6), we obtain the OP as

$$P_{out}(\gamma_{th}) = F_\gamma(\gamma_{th}). \quad (14)$$

## B. Higher-Order Amount of Fading

For the instantaneous SNR $\gamma$, the $n^{th}$-order amount of fading is defined as [26]

$$AF_\gamma^{(n)} = \frac{\mathbb{E}[\gamma^n]}{\mathbb{E}[\gamma]^n} - 1. \quad (15)$$

Substituting (13) in (15) yields to the $n^{th}$-order AF.

## C. Average BER

The average BER for a variety of binary modulations is introduced as [27, Eq.(12)]

$$\overline{P_b} = \frac{q^p}{2\,\Gamma(p)} \int_0^\infty \exp(-q\,\gamma)\, \gamma^{p-1}\, F_\gamma(\gamma)\, d\gamma, \quad (16)$$

where $p$ and $q$ are parameters that change for different modulation schemes [28]. Replacing $F_\gamma(\gamma)$ by its expression in (6) and utilizing [24, Eq.(07.34.21.0088.01)] with some algebraic manipulations, we obtain the BER as

$$\overline{P_b} = \frac{1}{2} - \frac{A\,q^p}{2\Gamma(p)} \sum_{k=0}^{m-1} \sum_{j=0}^{k} \frac{\left(\frac{m}{\Omega}\right)^{k-j}}{j!\,(k-j)!} \left(q + \frac{m}{\Omega}\right)^{j-k-p}$$

$$\times G_{r+1,3r+1}^{3r+1,1}\left[\frac{B\,m\,C}{\mu_r(q\,\Omega+m)} \,\middle|\, \begin{array}{c} 1-p-k-j, \kappa_1 \\ \kappa_2 \end{array}\right]. \quad (17)$$

For $m=1$, as a special case, we get the BER of the mixed Rayleigh/FSO fixed gain dual hop transmission systems with pointing errors given in [17, Eq.(11)]. At **high SNR** and similar to the CDF, the BER can be expressed asymptotically as

$$\overline{P_b} \underset{\mu_r \gg 1}{\approx} \frac{1}{2} - \frac{A\,q^p}{2\Gamma(p)} \sum_{k=0}^{m-1} \sum_{j=0}^{k} \frac{\left(\frac{m}{\Omega}\right)^{k-j}}{j!\,(k-j)!} \left(q + \frac{m}{\Omega}\right)^{j-k-p}$$

$$\times \sum_{i=1}^{3r+1} \left(\frac{\mu_r(q\,\Omega+m)}{B\,m\,C}\right)^{-\kappa_{2,i}}$$

$$\times \frac{\prod_{l=1;l\neq i}^{3r+1} \Gamma(\kappa_{2,l}-\kappa_{2,i})\Gamma(\kappa_{2,i}+p+k-j)}{\prod_{l=2}^{r+1}\Gamma(\kappa_{1,l}-\kappa_{2,i})}, \quad (18)$$

and can be further expressed via only the dominant term $j$.

## D. Ergodic Capacity

The ergodic capacity defined as $\overline{C} = \mathbb{E}[\log_2(1+\gamma)]$ can be written in terms of the CCDF of $\gamma$ as [29, Eq.(15)]

$$\overline{C} = 1/\ln(2) \int_0^\infty F_\gamma^c(\gamma)/(1+\gamma)\, d\gamma. \quad (19)$$

Using [30] to represent $(1+\gamma)^{-1}$ as $G_{1,1}^{1,1}\left[\gamma \,\middle|\, \begin{array}{c} 0 \\ 0 \end{array}\right]$, and utilizing the integral identity [27, Eq.(20)], we obtain the ergodic capacity in terms of the EGBMGF as

$$\overline{C} = \frac{A}{\ln(2)} \frac{\Omega}{m} \sum_{k=0}^{m-1} \sum_{j=0}^{k} \frac{1}{j!\,(k-j)!}$$

$$\times G_{1,0:1,1:r,3r+1}^{1,0:1,1:3r+1,0}\left[k-j+1\,\middle|\, \begin{array}{c} 0 \\ 0 \end{array}\,\middle|\, \begin{array}{c} \kappa_1 \\ \kappa_2 \end{array}\,\middle|\, \frac{\Omega}{m}, \frac{B\,C}{\mu_r}\right]. \quad (20)$$

An efficient Mathematica implementation of the EGBMGF is given in [27, Table II]. For $m=1$, as a special case, the ergodic capacity in (20) is in agreement with [17, Eq.(13)].

## V. NUMERICAL RESULTS

In this section, we present simulation and numerical results for different performance metrics of asymmetric dual-hop Nakagami-$m$-unified FSO relay transmission system with fixed gain relay, as an illustration of the analytical expressions given in the previous sections. The FSO link (i.e. the R-D link) is modeled as a unified Gamma-Gamma fading channel for weak ($\alpha = 2.902$ and $\beta = 2.51$), moderate ($\alpha = 2.296$ and $\beta = 1.822$), and strong ($\alpha = 2.064$ and $\beta = 1.342$) turbulent FSO channel conditions. In this section, the average SNR between the relay and the destination (R-D link) is set such that $\overline{\gamma_2} = 10$ dB except for the figures showing the asymptotic results where $\overline{\gamma_2}$ is varying. For the fixed gain scheme, the relay is set such us $C = 1$.

The outage probability performance for both heterodyne and IM/DD detection techniques versus the normalized average fading power of the RF link (i.e S-R link) is presented in Fig. 1. The effect of pointing error is fixed at $\xi = 1.1$. We can see from Fig. 1 that the analytical results provide a perfect match to the simulation results presented in this paper. It can also be observed that the heterodyne detection technique ($r = 1$) provides better performance than the IM/DD technique ($r = 2$). Moreover, it can be shown that the performance deteriorates as the atmospheric turbulence conditions get severe (i.e. the higher the values of $\alpha$ and $\beta$, the lower will be the OP) and vice versa.

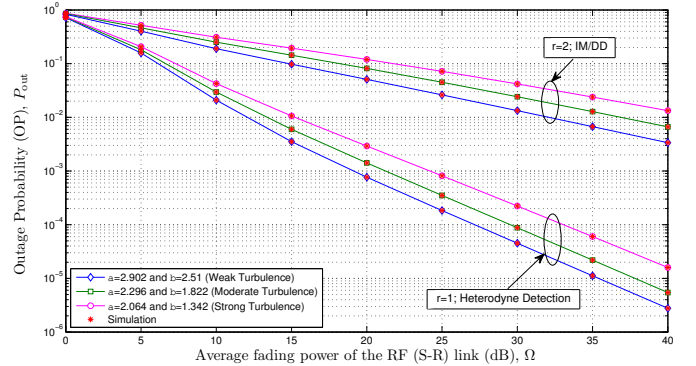

Fig. 1. OP showing the performance of both the detection techniques (heterodyne and IM/DD) under strong, moderate, and weak turbulent FSO channels for strong pointing error $\xi = 1.1$.

Fig. 2 presents the OP under both the detection techniques (heterodyne and IM/DD) for strong pointing error $\xi = 1.1$ along with the asymptotic results in high SNR regime. It can be shown that at high SNR, the asymptotic expression utilizing the Meijer's G function expansion and all the terms are considered in the summation in (7) converges quite fast to the exact result proving this asymptotic expression to be tight enough. Moreover, if we select the appropriate single dominant term, we get also a convergence to the exact result though relatively slower.

In Fig. 3, we illustrate the OP under IM/DD technique with varying effects of pointing error ($\xi = 1$ and $6.7$). As expected, the OP increases as the pointing error gets severe (i.e. the lower the values of $\xi$, the higher will be the OP). Additionally, it can be observed that for lower effect of the atmospheric turbulence, the respective performance gets better.

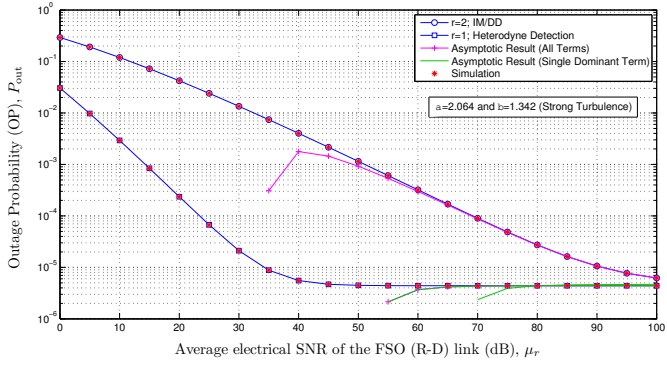

Fig. 2. OP showing the performance of both the detection techniques (heterodyne and IM/DD) under strong turbulence conditions for strong pointing error $\xi = 1.1$ along with the asymptotic results in high SNR regime for $\Omega = 20$ dB.

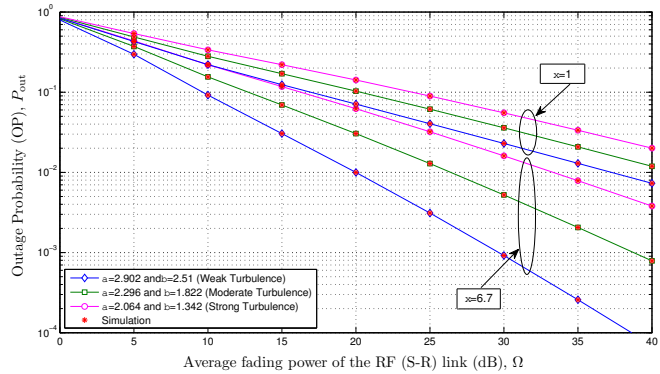

Fig. 3. OP showing the performance of IM/DD technique under strong, moderate, and weak turbulent FSO channels with varying effects of pointing error.

Fig. 4 demonstrates the average BER performance for differential binary phase shift keying (DBPSK) binary modulation scheme where $p = 1$ and $q = 1$ are the parameters of DBPSK, for both types of detection techniques (i.e. IM/DD and heterodyne) with fixed effect of the pointing error ($\xi = 1.1$). As clearly seen in the figure, the analytical results and the simulation results coincide. We can also see from this figure that the heterodyne detection technique outperforms the IM/DD technique. Moreover, it can be observed that the performance improves as the effect of the atmospheric turbulence drops.

Fig. 5 presents the average BER for DBPSK binary modulation scheme under IM/DD technique for varying effects of the pointing error ($\xi = 1$ and $6.7$) along with the asymptotic results in high SNR regime. It can be observed that at high SNR, the asymptotic expression utilizing the Meijer's G function expansion and considering all the terms in the summation in (18) converges quite fast to the exact result proving the tightness of this asymptotic approximation. Additionally, when we select the relevant single dominant term of (18) derived via Meijer's G function expansion, a slower convergence is clearly observed.

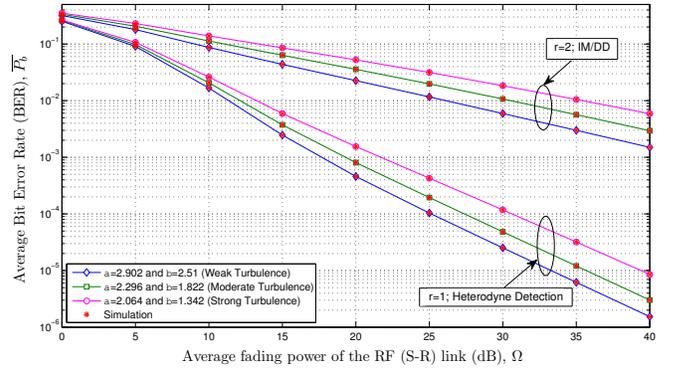

Fig. 4. Average BER of BDPSK binary modulation scheme showing the performance of both the detection techniques (heterodyne and IM/DD) under strong, moderate, and weak turbulent FSO channels for strong pointing error $\xi = 1.1$.

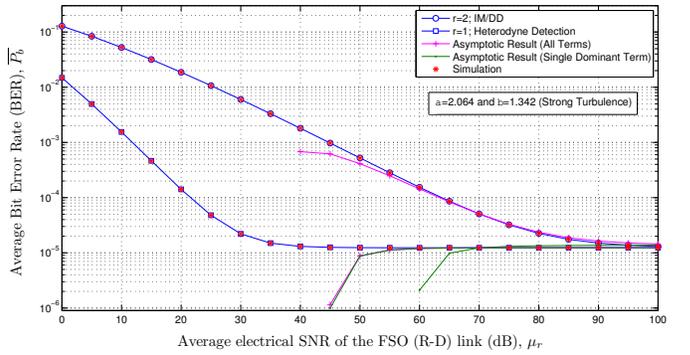

Fig. 5. Average BER of BDPSK binary modulation scheme showing the performance of both the detection techniques (heterodyne and IM/DD) under strong turbulence conditions for strong pointing error $\xi = 1.1$ along with the asymptotic results in high SNR regime for $\Omega = 20$ dB.

Fig. 6 depicts the average BER for DBPSK binary modulation scheme under IM/DD technique for varying effects of the pointing error ($\xi = 1$ and $6.7$). Expectedly, as the pointing error decreases ($\xi \to \infty$), the respective system performance gets better.

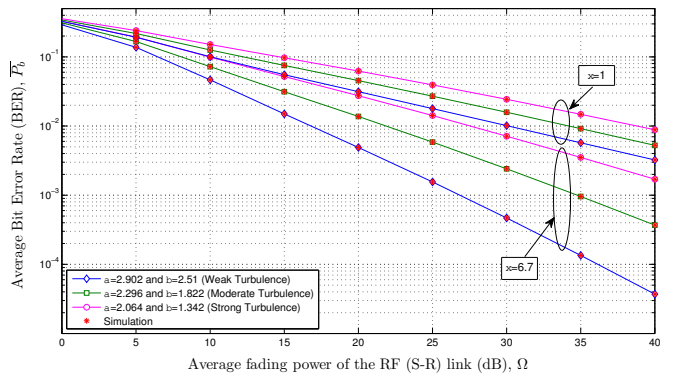

Fig. 6. Average BER of BDPSK binary modulation scheme showing the performance of IM/DD technique under strong, moderate, and weak turbulent FSO channels with varying effects of pointing error.

In Fig. 7, the ergodic capacity under both heterodyne and IM/DD detection techniques for varying effects of the pointing error ($\xi = 1$ and $6.7$) for strong turbulence conditions is presented. It can be observed that heterodyne detection performs much better than the IM/DD technique. Additionally, it can be shown that as the pointing error gets severe, the ergodic capacity decreases (i.e. the higher values of $\xi$, the higher will be the ergodic capacity).

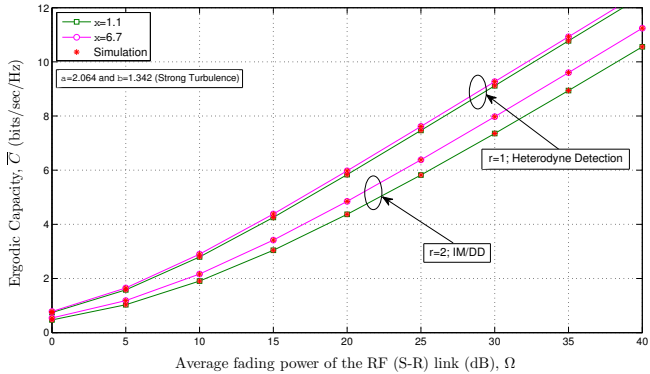

Fig. 7. Ergodic capacity results showing the performance of both heterodyne and IM/DD techniques under strong turbulence conditions for varying pointing errors.

## VI. Conclusion

In this work, we, for the first time, provided unified exact closed-form expressions for the PDF, the CDF, the MGF, and the moments of a dual-hop fixed gain relay system over the asymmetric links composed of both Nakagami-$m$ and unified Gamma-Gamma fading environments. From these formulas, we derived unified expressions for the higher-order AF, the average BER, and the ergodic capacity. In addition, we introduced asymptotic expressions at high SNR regime for the CDF, the MGF, the OP, and the average BER utilizing the Meijer's G function asymptotic expansion. We also demonstrated the impact of atmospheric turbulence conditions and pointing errors on the system performance.

## References

[1] W. Gappmair, "Further results on the capacity of free-space optical channels in turbulent atmosphere," *IET Communications*, vol. 5, no. 9, pp. 1262–1267, 2011.

[2] L. C. Andrews, R. L. Phillips, and C. Y. Hopen, *Laser Beam Scintillation with Applications*. SPIE Press, 2001.

[3] W. Popoola and Z. Ghassemlooy, "BPSK subcarrier intensity modulated free-space optical communications in atmospheric turbulence," *IEEE/OSA Journal of Lightwave Technology*, vol. 27, no. 8, pp. 967–973, Apr. 2009.

[4] J. Park, E. Lee, and G. Yoon, "Average bit-error rate of the alamouti scheme in Gamma-Gamma fading channels," *IEEE Photonics Technology Letters*, vol. 23, no. 4, pp. 269–271, Feb. 2011.

[5] M. Safari and M. Uysal, "Relay-assisted free-space optical communication," *IEEE Transactions on Wireless Communications*, vol. 7, no. 12, pp. 5441–5449, Dec. 2008.

[6] S. Navidpour, M. Uysal, and M. Kavehrad, "BER performance of free-space optical transmission with spatial diversity," *IEEE Transactions on Wireless Communications*, vol. 6, no. 8, pp. 2813–2819, Aug. 2007.

[7] S. Arnon, "Optimization of urban optical wireless communication systems," *IEEE Transactions on Wireless Communications*, vol. 2, no. 4, pp. 626–629, Jul. 2003.

[8] ——, "Effects of atmospheric turbulence and building sway on optical wireless communication systems," *Optics Letters*, vol. 28, no. 2, pp. 129–131, Jan. 2003.

[9] H. Sandalidis, T. Tsiftsis, G. Karagiannidis, and M. Uysal, "BER performance of FSO links over strong atmospheric turbulence channels with pointing errors," *IEEE Communications Letters*, vol. 12, no. 1, pp. 44–46, Jan. 2008.

[10] T. Tsiftsis, "Performance of heterodyne wireless optical communication systems over Gamma-Gamma atmospheric turbulence channels," *Electronics Letters*, vol. 44, no. 5, pp. 372–373, Feb. 2008.

[11] C. Liu, Y. Yao, Y. Sun, and X. Zhao, "Average capacity for heterodyne FSO communication systems over Gamma-Gamma turbulence channels with pointing errors," *Electronics Letters*, vol. 46, no. 12, pp. 851–853, Jun. 2010.

[12] M. Hasna and M.-S. Alouini, "A performance study of dual-hop transmissions with fixed gain relays," *IEEE Transactions on Wireless Communications*, vol. 3, no. 6, pp. 1963–1968, Nov. 2004.

[13] Y. Zhu, Y. Xin, and P.-Y. Kam, "Outage probability of Rician fading relay channels," *IEEE Transactions on Vehicular Technology*, vol. 57, no. 4, pp. 2648–2652, Jul. 2008.

[14] S. Datta, S. Chakrabarti, and R. Roy, "Error analysis of noncoherent FSK with variable gain relaying in dual-hop Nakagami-m relay fading channel," in *Proc. 2010 International Conference on Signal Processing and Communications (SPCOM'2010)*, Jul. 2010, pp. 1–5.

[15] G. Karagiannidis, "Performance bounds of multihop wireless communications with blind relays over generalized fading channels," *IEEE Transactions on Wireless Communications*, vol. 5, no. 3, pp. 498–503, Mar. 2006.

[16] E. Lee, J. Park, D. Han, and G. Yoon, "Performance analysis of the asymmetric dual-hop relay transmission with mixed RF/FSO links," *IEEE Photonics Technology Letters*, vol. 23, no. 21, pp. 1642–1644, Nov. 2011.

[17] I. S. Ansari, F. Yilmaz, and M.-S. Alouini, "On the performance of hybrid RF and RF/FSO fixed gain dual-hop transmission systems," in *Proceedings of The Second Saudi International Electronics, Communications and Photonics Conference (SIECPC' 2013)*, Riyadh, Saudi Arabia, Apr. 2013, pp. 1–6.

[18] ——, "On the performance of mixed RF/FSO variable gain dual-hop transmission systems with pointing errors," in *Proceedings of IEEE 78th Vehicular Technology Conference (VTC Fall' 2013)*, Las Vegas, USA, Sep. 2013, pp. 1–6.

[19] ——, "Impact of pointing errors on the performance of mixed RF/FSO dual-hop transmission systems," *IEEE Wireless Communications Letters*, vol. 2, no. 3, pp. 351–354, Jun. 2013.

[20] M. Simon and M.-S. Alouini, "Digital communication over fading channels," *IEEE Transactions on Wireless Communications*, 2005.

[21] I. S. Gradshteyn and I. M. Ryzhik, *Table of Integrals, Series and Products*. New York: Academic Press, 2000.

[22] I. S. Ansari, F. Yilmaz, and M.-S. Alouini, "A unified performance of free-space optical links over Gamma-Gamma turbulence channels with pointing errors," submitted to *IEEE Transactions on Communications*, technical report available at http://hdl.handle.net/10754/305353.

[23] H. Sandalidis, T. Tsiftsis, and G. Karagiannidis, "Optical wireless communications with heterodyne detection over turbulence channels with pointing errors," *Journal of Lightwave Technology*, vol. 27, no. 20, pp. 4440–4445, Oct. 2009.

[24] I. Wolfram Research, *Mathematica Edition: Version 8.0*. Champaign, Illinois: Wolfram Research, Inc., 2010.

[25] M. D. Springer, *The Algebra of Random Variables (Probability & Mathematical Statistics)*. John Wiley & Sons Inc, 1979.

[26] F. Yilmaz and M.-S. Alouini, "Novel asymptotic results on the high-order statistics of the channel capacity over generalized fading channels," in *IEEE 13th International Workshop on Signal Processing Advances in Wireless Communications (SPAWC'2012)*, 2012, pp. 389–393.

[27] I. S. Ansari, S. Al-Ahmadi, F. Yilmaz, M.-S. Alouini, and H. Yanikomeroglu, "A new formula for the BER of binary modulations with dual-branch selection over generalized-K composite fading channels," *IEEE Transactions on Communications*, vol. 59, no. 10, pp. 2654–2658, Oct. 2011.

[28] N. Sagias, D. Zogas, and G. Karagiannidis, "Selection diversity receivers over nonidentical Weibull fading channels," *IEEE Transactions on Vehicular Technology*, vol. 54, no. 6, pp. 2146–2151, Nov. 2005.

[29] A. Annamalai, R. Palat, and J. Matyjas, "Estimating ergodic capacity of cooperative analog relaying under different adaptive source transmission techniques," in *Proceedings of 2010 IEEE Sarnoff Symposium*, Apr. 2010, pp. 1–5.

[30] A. Mathai and R. Saxena, *The H-function with applications in statistics and other disciplines*. New York: Wiley Eastern, 1978.